\documentclass[10pt,preprint2]{aastex}
\usepackage{graphicx}
\usepackage{bm}
\usepackage{natbib}
\usepackage{textcomp}
\usepackage{amsmath}
\usepackage{amssymb}

\shorttitle{CLV modeling of Ca {\sc i} 4227 \AA{} line}
\shortauthors{Supriya et al.}

\begin{document}
\title{Center to limb observations and modeling of the Ca {\sc i} 4227 \AA{} line}

\author{H. D. Supriya$^{1}$, H. N. Smitha$^{1}$, K. N. Nagendra$^{1}$, 
J. O. Stenflo$^{2,3}$, M. Bianda$^{3}$, R. Ramelli$^{3}$, B. Ravindra$^{1}$, and L. S. Anusha$^{4}$}
\affil{$^1$Indian Institute of Astrophysics, Bangalore 560034, India}
\affil{$^2$Institute of Astronomy, ETH Zurich, CH-8093 \  Zurich, Switzerland }
\affil{$^3$Istituto Ricerche Solari Locarno, Via Patocchi, CH-6605 Locarno-Monti, Switzerland}
\affil{$^4$Max Planck Institut f\"ur Sonnesystemforschung, Justus-von-Liebig-Weg 3, D-37077 G\"ottingen, Germany}

\email{hdsupriya@iiap.res.in; smithahn@iiap.res.in; knn@iiap.res.in; stenflo@astro.phys.ethz.ch; 
mbianda@irsol.ch; ramelli@irsol.ch; ravindra@iiap.res.in; bhasari@mps.mpg.de}

\begin{abstract}
The observed center-to-limb variation (CLV) of the scattering polarization in 
different lines of the Second Solar Spectrum can be used to constrain the height 
variation of various atmospheric parameters, in particular the 
magnetic fields via the Hanle effect. Here we attempt to model non-magnetic CLV observations 
of the $Q/I$ profiles of the Ca\,{\sc i}~4227\,\AA\ line recorded with the ZIMPOL-3 at IRSOL. 
For modeling, we use the 
polarized radiative transfer with partial frequency redistribution 
with a number of realistic 1-D model atmospheres. We find that all 
the standard FAL model atmospheres, used by us, fail to simultaneously fit 
the observed ($I$,\,$Q/I$) at all the limb distances ($\mu$). 
However, an attempt is made to find a single model 
which can provide a fit at least to the CLV of the observed $Q/I$ instead 
of a simultaneous fit to the ($I$,\,$Q/I$) at all $\mu$. 
To this end we construct a new 1-D model by combining two of the standard models 
after modifying their temperature structures in the appropriate height ranges.  
This new combined model closely reproduces the 
observed $Q/I$ at all the $\mu$, but 
fails to reproduce the observed rest intensity at 
different $\mu$. Hence we find that no single 1-D model atmosphere succeeds in providing a 
good representation of the real Sun. 
This failure of 1-D models does not however cause an impediment to the 
magnetic field diagnostic potential of the Ca\,{\sc i}~4227\,\AA\ line. 
To demonstrate this we deduce the field strength at various $\mu$ positions without 
invoking the use of radiative transfer. 
\end{abstract}


\maketitle

\section{Introduction}
\label{intro}
Coherent scattering processes in the solar atmosphere causes the emitted radiation  
to be linearly polarized. The spectrum so obtained is referred to as the 
Second Solar Spectrum (SSS). This linearly polarized spectrum gets modified 
at the line core in the presence of weak magnetic fields due to the Hanle effect. One 
of the commonly observed and well studied lines in the SSS is the 
strong, chromospheric Ca {\sc i} 4227 \AA{} line \citep[see for example,][]{stenetal80, sten82,
gandorfer2002}. 
This is a normal Zeeman triplet line arising due to transition between 
the atomic states with total angular momentum $J=0 \to 1 \to 0$. It 
exhibits largest scattering polarization among all lines in the 
Sun's visible spectrum \citep{stenetal80}. 
The core of the Ca {\sc i} 4227 \AA{} line is formed around a height of about 
1000 km above the photosphere making it chromospheric in nature.
The Hanle effect in the core of the Ca {\sc i} 4227 \AA{} line 
was first observed by \citet{sten82}. 

One of the interesting feature exhibited by this line is 
the spatial variation of the wing polarization in $(Q/I, U/I)$ spectra along the 
spectrograph slit. \citet{biandaetal03} noticed these variations 
in the observations near an active region. Later these spatial 
variations were also observed in quiet regions by 
\citet{sampetal09} and were interpreted to be 
arising due to the local inhomogeneities in the atmospheric layers. A detailed modeling of the 
Ca {\sc i} 4227 \AA{} line profile observed in a quite region near the solar limb is presented 
by \citet{lsa10}. The authors employed last scattering approximation to model them. 
Observations of the forward scattering Hanle effect in the Ca {\sc i} 4227 \AA{} line near the disk 
center were performed by \citet{bianda11}. Subsequently these observations were 
modeled by \citet{FS11} to determine the chromospheric weak magnetic fields. 

One of the early attempts to model the polarization profiles 
of the Ca~{\sc i}~4227~\AA\ line
observed in both quiet and active regions was by \citet{mf92}.
The author used the observations of \citet{sten82}.
Her treatment included the effects of partial frequency redistribution (PRD)
and radiative transfer (RT). 

To better understand the physics of scattering and to exploit 
it for various diagnostic purposes, 
we need to systematically study the center-to-limb variation (CLV) of the SSS. 
This will help us sample the height information of the atmospheric parameters 
and magnetic fields, as observations made at different lines of sight (LOS) sample different heights in the solar atmosphere.
Few attempts have been made so far in detailed modeling of the CLV observations of 
($I$, $Q/I$) spectra of atomic and molecular lines in the SSS. 
The most challenging aspect of such CLV modeling is to find a single model 
atmosphere which can fit both $I$ and $Q/I$ at all limb distances $\mu~(=\cos \theta)$,
simultaneously. One such attempt was made by \citet{holandsten07} 
to model CLV of the Ca {\sc ii} K line. They discuss the possibility of constructing a 
two-component atmospheric model (using a combination of a hot and a cool atmospheric component) to 
achieve a fit to the observed CLV profiles. A height dependent mixing ratio was 
required by them in order to simultaneously fit $I$ and $Q/I$ spectra at all the limb distances. 
They also demonstrate that a single atmospheric model with optimized temperature structure
can be used to achieve a fit to the Ca {\sc ii} K line CLV data. 
However they find that different extents of modification in 
the temperature structure are required for different limb distances ($\mu$). 
Another paper in which such CLV studies 
have been done is that of \citet{shapiro11} who consider the molecular CN violet system. 
They discuss the general problems involved in obtaining a 
simultaneous fit to the $I$ and $Q/I$ profiles using the standard  
1-D single atmospheric model, as well as two-component atmospheric models. 
They finally construct an anisotropy modified single 1-D 
atmospheric model to simultaneously fit $I$ and $Q/I$ at all $\mu$.

The CLV of the Ca~{\sc i} 4227~\AA\ line away from the active regions was 
first observed by \citet{stenetal80} and analyzed by \citet{aueretal80}.
The CLV of the line center polarization observed by \citet{stenetal80}
was later used by \citet{mf94} to study the Hanle effect 
due to the magnetic field canopies in the chromosphere. 
The CLV observations of this line was also done by \citet{bianda98,bianda99}. 
In this paper we attempt a detailed simultaneous modeling of the observed CLV of both $I$
and $Q/I$ profiles of the Ca {\sc i} 4227 \AA{} line. 
For this purpose we solve the polarized RT equation by taking account of 
PRD effects in the non-magnetic regime. Standard 1-D atmospheric FAL models 
\citep{font93, avrett95} are used to obtain a fit to the $(I, Q/I)$ spectra. We find that it is not 
possible to achieve a simultaneous ($I, Q/I$) fit to the CLV observations with a single 1-D model atmosphere. 
If we consider the CLV of $Q/I$ alone, then we find it necessary to modify the original 
temperature structure of the standard FAL atmospheric models to obtain a fit. 
Such modifications of the original temperature structure were also used in 
previous works by \citet{holandsten07, smietal12, smietal13}. In the present paper, the original 
temperature structure of the standard FAL-A atmosphere is modified. Later, the 
modified FAL-A ($\overline{{\rm FALA}}$) is combined with FAL-X atmospheric 
model to construct a single component model. It turns out that this newly constructed 
combined model can closely reproduce the observed $Q/I$ at different limb distances. 

In Section~{\ref{obsevations} we give the details of the CLV observations. Section~{\ref{modeling}} 
is devoted to the modeling procedure and the results. Section~\ref{depth-dependence} describes the 
observational analysis to determine the magnetic fields. Concluding 
remarks are given in Section~{\ref{conclusion}}.

\section{Observational details}
\label{obsevations}
\begin{figure*}
\centering
\includegraphics[scale=0.6]{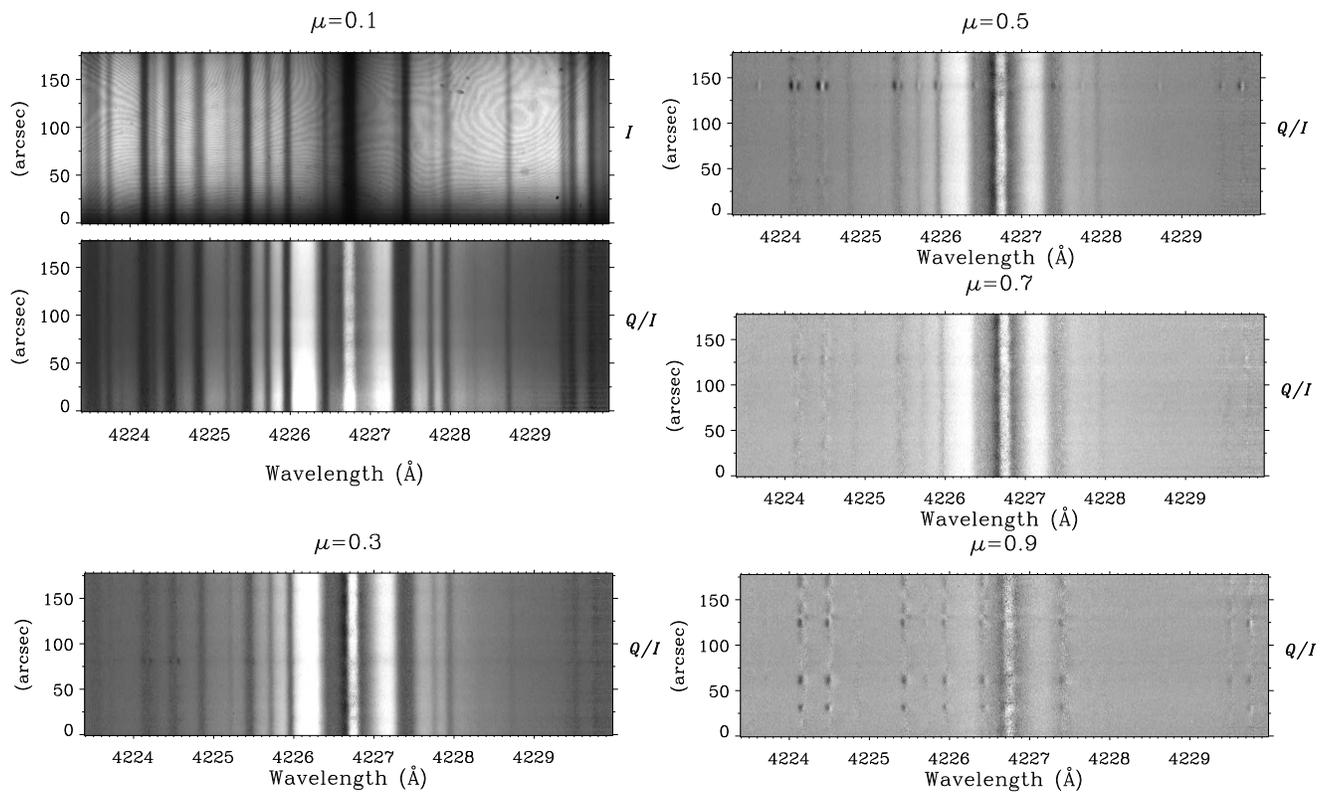}
\caption{CCD images of $I$ (only for $\mu = 0.1$) and $Q/I$  
at 5 selected $\mu$ values of the Ca {\sc i} 4227 \AA{} line.
The observations were taken on October 16, 2012 at IRSOL in Switzerland.}
\label{ccd}
\end{figure*}

The CLV observations of the Ca {\sc i} 4227 \AA{} line presented in this paper were
obtained with the Zurich Imaging Polarimeter-3 (ZIMPOL-3) \citep{ramellietal10} 
at IRSOL (Istituto Ricerche Solari Locarno) in Switzerland on October 16, 2012. 
The observations were taken at 14 different $\mu$ positions (0.10, 0.15, 0.20, 0.25, 0.30, 0.35, 
0.40, 0.45, 0.50, 0.60, 0.70, 0.80, 0.90 and 1.0) starting from the
heliographic north pole at $\mu=0.1$ up to the disk center at $\mu=1$. Figure~\ref{ccd} 
shows the CCD images of the Ca {\sc i} 4227 \AA{} line at 5 selected positions on 
the solar disk. The polarization modulation was done with a piezo-elastic modulator
(PEM). The spectrograph slit was 60 $\mu$m wide corresponding to
a spatial extent of 0.5$\arcsec$ on the solar disk. The CCD covered
190$\arcsec$ along the slit. The CCD images have 140 
effective pixel resolution elements in the spatial 
direction, with each element corresponding to 1.38$\arcsec$, and 
1240 pixels in the wavelength direction, with one pixel corresponding to 5.30 m\AA{}.

To keep the solar image position stable, the primary image guiding system is 
used \citep{kulvere11}. In addition, below $\mu = 0.35$ a rotating 
glass tiltplate is used to keep under control the distance between the 
spectrograph slit and the solar limb image. The slit jaw image is digitized 
by a dedicated CCD camera. An algorithm recognizes the solar limb and 
the spectrograph slit position on the image. This allows the calculated desired distance between the 
limb and the slit to be set by automatic control of the tilt plate. Note 
that this plate is set after the polarization analyzer and hence does not 
introduce spurious polarization signatures.
At each $\mu$ position a measurement is obtained by adding 300 single frames, 
each of them are obtained with an exposure time of 1 sec. 
Therefore the effective integration time is 8 minutes. The 
precision of the pointing at a chosen $\mu$ position over 8 minutes using 
the tiltplate is limited to about 1$\arcsec$, which is less than the size of one pixel.

An improvement of these measurements is related to the absolute precision 
which we could reach in measuring $Q/I$. Previously, the zero polarization value 
needed to be defined manually was based on indirect considerations (for example 
the CLV of the continuum polarization). For the data set described in this 
paper we could reach an absolute precision of about $5\times10^{-5}$.  This is 
mainly due to: a) the precise control of limb distance, allowed by the 
tiltplate system described above, b) the improved control of the rotation 
of the optical devices, including the polarization analyzer in front of the 
slit (to compensate for the image rotation originated by the Gregory Coud\`e 
telescope), and c) the optical compensation of the instrumental linear 
polarization, which is a source of variable offset effects, with an oriented glass 
plate set in front of the polarization analyzer.
It was thus possible to subtract from every measurement done at a defined 
$\mu$ position, the polarization level measured at disk center in a quiet 
region. For symmetry reasons the linear polarization in the continuum 
is expected to be zero at the disk center. In this way all residual instrumental linear 
polarization signatures are taken into account.

\subsection{Stray light correction}
\label{stray-light}
The observed profiles contain contribution from the spectrograph stray light which is about 2 \% of the 
continuum intensity. Here we correct both the intensity and polarization profiles for stray light. 
The effect of stray light, including both its intensity and polarization, was treated in \citet{stenflo74}. 
Below are the details of the procedure we have followed.

In the absence of stray light but with instrumental polarization (cross talk from $I(\lambda)$), 
the $I(\lambda)$ and $Q(\lambda)$ parameters after polarization calibration are 
$I'(\lambda)$ = $I(\lambda)$ and $Q'(\lambda)$ = $Q(\lambda) + M_{21}I(\lambda)$, if
we assume that the Mueller matrix has been normalized ($M_{11} = 1$), there is no telescope 
depolarization or it has been calibrated away ($M_{22} = 1$), and that polarization cross talk 
from $U$ and $V$ can be disregarded. $M_{21}$ is the spectrally flat instrumental polarization, 
which for convenience will be renamed as $p_z$, since it represents a flat offset of the zero point 
of the polarization scale.

In the presence of stray light with an intensity that is a fraction $s$ of the continuum 
intensity $I_c$ and has a polarization $p_s$, the apparent or observed Stokes parameters are
\begin{eqnarray}
 && \!\!\!\!\!\!\!\!\!\!\!I_{obs}(\lambda) = I(\lambda) + sI_c, \nonumber \\&&
\!\!\!\!\!\!\!\!\!\!\!Q_{obs}(\lambda) = Q(\lambda) + p_z (I(\lambda) + sI_c) + p_s sI_c.
\label{obsestokes}
\end{eqnarray}
We now introduce the notation $r(\lambda)$ = $I(\lambda)/I_c$ and 
$r_{obs}(\lambda)$ = $I_{obs}(\lambda)/I_{obs,c}$ for the rest intensities. 
For clarity we attach a 
$\lambda$ to the quantities that are spectrally structured, in contrast to the three free parameters of 
our problem, namely, $s$, $p_s$, and $p_z$, which are constant and spectrally flat. Then
\begin{equation}
r(\lambda) = (1 + s) r_{obs}(\lambda) - s.
\label{rlambda}
\end{equation}
Similarly we define the intrinsic polarization $p({\lambda}) = Q(\lambda)/I(\lambda)$ 
and the apparent polarization $p_{obs}(\lambda)$ = $Q_{obs}(\lambda)/I_{obs}(\lambda)$. 
One can easily show that
\begin{equation}
 p({\lambda}) = \bigg(1 + \frac{s}{r(\lambda)}\bigg) [ p_{obs}(\lambda) - p_z ] - 
\frac{s}{r(\lambda)} p_s.
\label{plambda}
\end{equation}
To calculate $p(\lambda)$ from the observations we need to insert the expression for $r(\lambda)$ 
from Equation~(\ref{rlambda}) into Equation~(\ref{plambda}).

$p_s$ represents the intrinsic polarization as averaged over the wide spectral range 
that contributes to the stray light. A major source of spectrograph stray light are grating 
ghosts that sample discrete wavelengths spread over a large wavelength range. In the 
absence of other information, the best estimate of $p_s$ is probably $p_s \approx p_c$, 
i.e., to set it equal to the continuum polarization.

The determination of $p_z$ is best made for a disk center recording that is done immediately 
before or after the measurement at the given $\mu$ position (so that one can assume 
that the instrumental polarization has not changed). At disk center the solar scattering 
polarization is zero, so the apparent polarization that we see is simply $p_z$ (in contrast to 
measurements of disk center, where $p_z$ is mixed with intrinsic solar polarization).

It is important to realize that the problem of correcting for the zero point of the 
polarization scale is entirely decoupled from the stray light issue. It is the first step to be 
done, and it gives us the spectrum 
\begin{equation}
 p'(\lambda) = p_{obs}(\lambda) - p_z,
\label{pprimelam}
\end{equation}
which would equal to $p({\lambda})$ in the absence of stray light. To correct $p'(\lambda)$ 
for stray light we do not need to refer to $p_z$ or disk center observations. 
The way in which the stray light correction enters can be seen by rewriting Equation~(\ref{plambda}) 
as
\begin{equation}
 p(\lambda) = p'(\lambda) + \frac{s}{r(\lambda)} (p'(\lambda) - p_s).
\label{plamnew}
\end{equation}
If the stray light were unpolarized, then the stray light scaling factor $s/{r(\lambda)}$, 
which is large where the rest intensity $r(\lambda)$ is low, acts to amplify the polarization 
amplitudes $p'(\lambda)$. In the presence of stray light polarization, however, 
the scaling factor only acts on the amplitude 
with respect to the $p_s$ level rather than with respect to the zero level. 
The stray light polarization therefore reduces the effect of the stray light correction. 
For polarization amplitudes that is equal to $p_s$ (which represents a broad band polarization 
background that may be approximated with the continuum polarization level $p_c$), the
stray light correction does not have any effect at all. For the Ca {\sc i} 4227\AA{} line, however, the
core polarization is usually larger than $p_c$. In this particular case the stray light polarization
becomes a second-order effect (since $sp_c$ is a product of two small quantities).
Parameter $s$ is determined exclusively from fitting the Fourier Transform Spectrum (FTS) 
of \citet{kurucz84}, in
the same context as the spectral broadening is determined. 
The above considerations give us a rather well defined procedure to determine (within
the framework of our idealized model) unique estimates of the parameters $s$, $p_z$, and $p_s$.
Using these estimates we can correct both the $I(\lambda)$ and 
$Q(\lambda)/I(\lambda)$ spectra 
for stray light. In other sections we have dropped the $\lambda$ dependence 
of $I$ and $Q$ for notational simplicity.

While the observed spectra are corrected for stray light, they have not been 
corrected for spectral broadening, because the instrumental profile is not known with 
the precision that is needed to allow a deconvolution. 
The theoretical spectra on the other hand, which are used for comparison with the 
observed spectra, need to be spectrally broadened to emulate the observations. However, 
one should not apply stray light to the theoretical spectra, since one can easily do the 
correction to the observed spectra itself. In this way we keep the presentation of the 
theoretical results independent of the particular properties of the instrument used for the 
observations, with the single exception of spectral broadening.

\begin{figure*}
\centering
\includegraphics[scale=0.7]{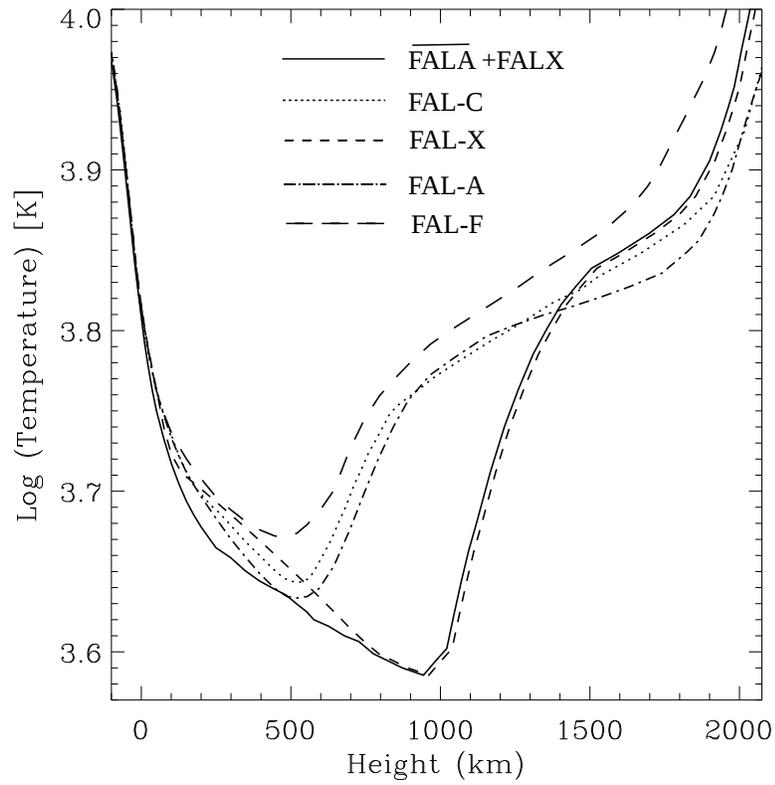}
\caption{Temperature structure of four standard models FAL-A, FAL-C, FAL-F and FAL-X used in our 
studies. Along with these standard models the temperature structure of the new model atmosphere 
$\overline{{\rm FALA}} + {{\rm FALX}}$ is also shown.}
\label{temp-structure}
\end{figure*}

\begin{figure*}
\centering
\includegraphics[scale=0.5]{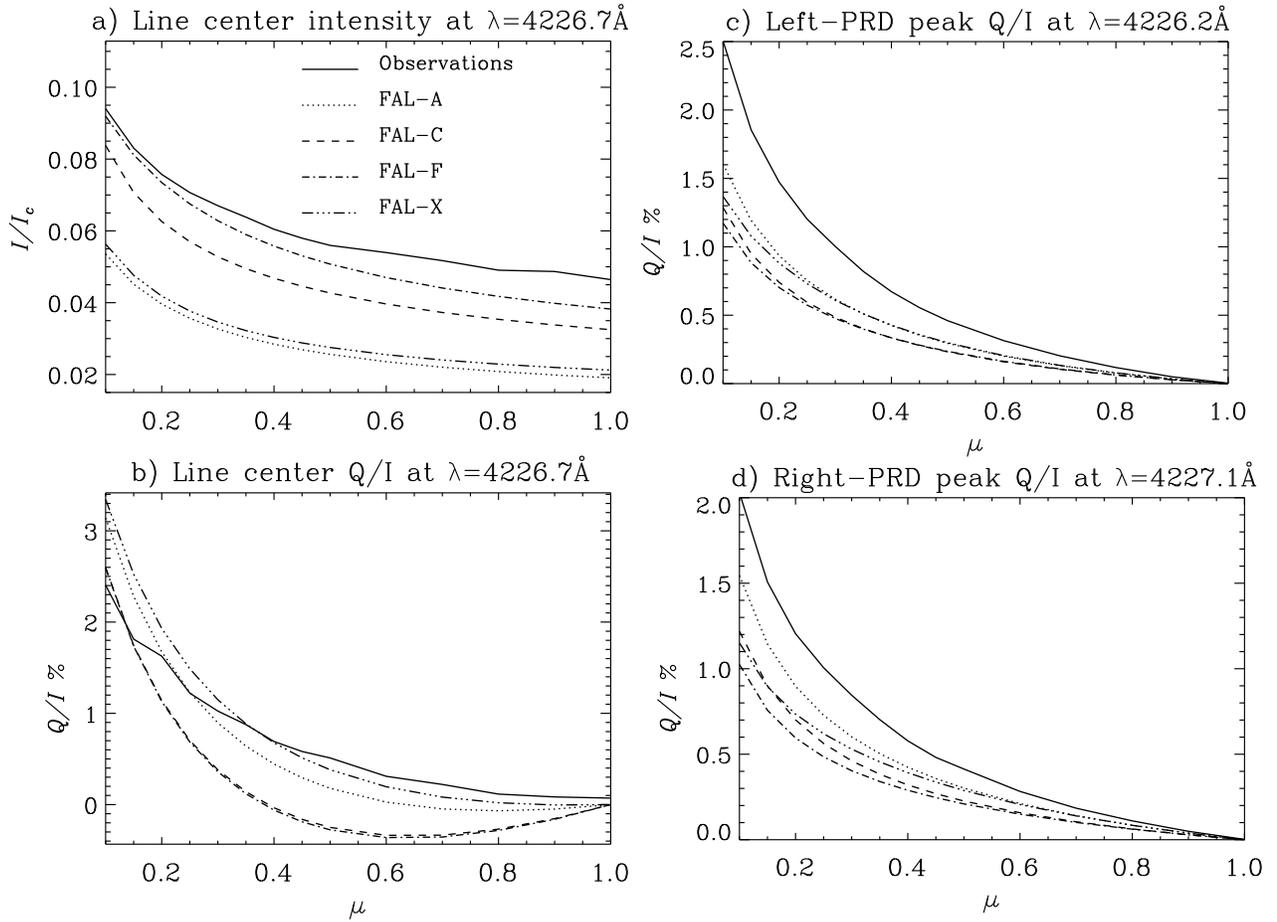}
\caption{Observed (solid line) and calculated intensity and polarization signals as function 
of $\mu$ (observed at 14 points) at three different wavelength positions in the line profile. 
The spectra are calculated for standard models FAL-X, FAL-C, FAL-A, and FAL-F.}
\label{clv-plots}
\end{figure*}

\section{Modeling of the CLV observations}
\label{modeling}
In this section we describe the modeling procedure we have followed to model the 
CLV of the Ca~{\sc i} 4227~\AA\ line. We started with an aim of finding a single 1-D atmosphere 
that can simultaneously 
fit ($I$, $Q/I$) CLV of the Ca~{\sc i} 4227~\AA\ line. 
As a first step towards this, we 
begin the modeling of the observed profiles by 
solving the polarized RT equation for a two-level atom.
The one-dimensional polarized RT equation along with the other necessary equations,
in the non-magnetic regime, used in this paper are described 
in detail elsewhere \citep[see][]{lsa10}. The elastic collision rates used in this paper 
are computed following the theory presented in \citet{barklem97}. 
Also, the modeling is done here using a two-stage process 
where the intensities are computed using a PRD capable MALI 
(Multi-level Approximate Lambda Iteration) code 
of \citet{rh01} in the first stage. In the second stage, 
the polarization profiles are computed perturbatively by solving the polarized transfer 
equation. The details of such a two-stage 
modeling procedure is described in \citet[][see also \citealt{lsa10}]{reneetal05}.
The atom model of the Ca~{\sc i} used in the present paper is same as the one
discussed in \citet{lsa10}. Hence we do not repeat the details here.

In our studies we use four standard 1-D model atmospheres of the Sun 
namely - FAL-A, FAL-C, FAL-F \citep{font93} and FAL-X \citep{avrett95}. The temperature structure 
of these models are shown in Figure ~\ref{temp-structure}. Along with the standard models, the 
temperature structure of our newly constructed model $\overline{{\rm FALA}} + {{\rm FALX}}$ is 
also shown in the figure, details of which will 
be discussed in Section~\ref{newmodel}.

\subsection{CLV behavior at three wavelength positions}
The observed $Q/I$ profiles of the Ca~{\sc i}~4227~\AA\ line show 
three prominent features. They are, the line center at 4226.7~\AA, the 
blue wing PRD peak at 4226.2~\AA\ and the red wing PRD peak at 4227.1~\AA. 
While the line center of the Ca {\sc i} 4227 \AA{} line is formed within a height range of 
700 - 1000 km (covering $0.9 \leqslant \mu \leqslant 0.1$), the blue and the red wing PRD peaks are formed 
at a height of 150-250 km above 
the level where the vertical continuum optical depth at 5000 \AA{} is unity. 
To get an idea of the behavior of the polarized spectra as a function of $\mu$, we plot 
the angular dependence of intensity (only at line center) and linear polarization at these chosen 
wavelength positions in Figure~\ref{clv-plots} computed using the standard 1-D model atmospheres. 
As expected, it can be seen from Figure~\ref{clv-plots} that the
degree of linear polarization decreases to zero towards the disk 
center due to symmetry in the scattering geometry.

Panels (a) and (b) of Figure~\ref{clv-plots} show a comparison of the observed and theoretical 
CLV of $I$ and $Q/I$ respectively at the line center wavelength. 
We see that the hottest model FAL-F (dot-dashed line) is more suited for modeling the CLV of 
line center intensity.
But the same model is not at all good for $Q/I$. Instead it is the coolest model FAL-X which provides 
the closest fit to the observed $Q/I$. 
This contrasting behavior seems to point at the fact that we need two different temperatures 
to simultaneously fit $I$ and $Q/I$ at the line center. From Figure~\ref{clv-plots} (b) 
we see that the theoretical profile computed using FAL-A also falls close to the observed 
CLV profile of $Q/I$. 
We consider the FAL-X atmosphere rather than FAL-A to provide a better fit to the 
observed $Q/I$ at the line center for the following reasons:  
At the line center and when $\mu$ is small, we always expect the non-magnetic $Q/I$ amplitude to be 
larger as compared with the observed $Q/I$, because the observed $Q/I$ 
includes depolarization by magnetic fields. As we go to larger $\mu$ the magnetic fields 
may enhance the core polarization amplitude. Such a behavior was noted 
by \citet{mf94}, who points out that there are enhancement effects due to 
magnetic fields when $\mu > 0.4$ (this will be discussed in detail in Section~\ref{results}). 
We found the FAL-X model to better satisfy this behavior. 
Thus we see from Figure~\ref{clv-plots} (b) that the theoretical 
values of $Q/I$ at line center when computed with FAL-X 
are larger than the observed $Q/I$ for $\mu < 0.4$, while FAL-A shows this behavior 
only for $\mu < 0.25$. 
Besides this, the theoretical $Q/I$ computed with FAL-A falls much below the 
observed $Q/I$ as we move towards 
larger values of $\mu$. 
For these reasons we consider FAL-X to give a consistent overall fit to the observed CLV 
of $Q/I$ at line center, while FAL-A does not. 

On the other hand, Figures~\ref{clv-plots} (c) 
and (d) show the CLV profiles of $Q/I$ at the blue and red wing PRD peak wavelength positions 
respectively. We notice that both FAL-F and FAL-X model atmospheres fail to provide a fit to the 
PRD peaks. It is the theoretical CLV profiles from the FAL-A model that falls closest to the 
observed CLV of $Q/I$. 
Thus we do not find a single 1-D atmospheric model which can 
provide a fit to the entire Stokes ($I$, $Q/I$) profiles simultaneously. As a next step we 
explore the possibility of obtaining a fit to the CLV of the Stokes profiles through small 
modification of the temperature structure at appropriate heights, of the original FAL models.  

\subsection{Theoretical fit to the CLV of $Q/I$ profiles}
\label{newmodel}
From Figures~\ref{clv-plots} (c) and (d) we see that though the theoretical profiles from the 
model FAL-A falls closest to the observed CLV of $Q/I$, it fails to provide a 
satisfactory fit to the {$Q/I$ observations}. In order to obtain a better fit to the 
observed CLV of $Q/I$ profiles, we adopt 
modification of temperature structure 
at the heights where PRD peaks are formed. 
We focus our attention only on fit to the $Q/I$ profiles. 
Since FAL-A model atmosphere provides closest fit 
to the observed $Q/I$ profiles, we choose this model for further modifications. 
Accordingly, the temperature of the FAL-A standard model at these heights is reduced by about 200 K. 
This newly constructed model is denoted as $\overline{{\rm FALA}}$. 
This new model provides a better fit to the wing PRD peaks at all the limb distances. 

After achieving a fit to the PRD peaks, we concentrate on obtaining a fit to the 
linear polarization at the line center. From Figure~\ref{clv-plots} (b) we see that it is the 
FAL-X model atmosphere which fits the observed profiles, the closest. Thus, 
to obtain a satisfactory fit to the entire $Q/I$ profile 
we need to combine these two model atmospheres ($\overline{{\rm FALA}}$ and FAL-X) at appropriate 
heights.
The two models are combined such that the new model atmosphere has the temperature structure of 
$\overline{{\rm FALA}}$ up to a height of ~400 km, and the temperature structure 
of FAL-X 
in the heights above 400 km. 
The temperature structure of the new combined model atmosphere 
($\overline{{\rm FALA}} + {{\rm FALX}}$) 
is shown as the solid line in Figure~\ref{temp-structure}.
The results obtained using this combined model atmosphere are discussed below.

\subsubsection{Results from the new combined model atmosphere}
\label{results}
The theoretical profiles obtained using $\overline{{\rm FALA}} + {{\rm FALX}}$ model atmosphere 
are shown in Figure~\ref{profiles} (dotted line). Also a comparison between the observed and theoretical 
$Q/I$ CLV curves at the blue and red wing PRD peaks and at the line center wavelength 
using the combined model is shown in Figure~\ref{clv-new} (dotted line).
These theoretical profiles show that we obtain an overall satisfactory fit 
to the $Q/I$ profiles at all the $\mu$ positions using the 
combined model atmosphere. The theoretical profiles computed using the combined model atmosphere 
in Figures~\ref{profiles} and \ref{clv-new} include suitable spectral smearing.  
This is done by convolving the theoretical spectra with a Gaussian profile having a 
FWHM of 50 m\AA{}. 
The smearing accounts for both the instrumental broadening (40 m\AA{}) and 
the broadening by macro-turbulent velocity fields (30 m\AA{}). The macro-turbulent 
smearing corresponds to a velocity of 1.28 km/s. For the deep lines like the 
Ca {\sc i} 4227 \AA{}, the effects of the stray light corrections (the 
stray light correction procedure is described in 
Section~\ref{stray-light}) are much more important than the smearing. The intensity ($I$) 
profiles from the combined model seem to fit the observed data at all the $\mu$ 
positions, with the exception of the line core, where we fail to get a satisfactory fit. At the 
formation heights of the line center, a rest intensity fit requires a 
hotter atmospheric model like the FAL-F which 
is not suitable for achieving a good fit to the 
$Q/I$ profile - which indeed requires cooler models like the FAL-X. 
In spite of the carefully determined stray light correction (by $s$ = 2\%) to the 
observed Stokes $I$ profiles, the central line depth still does not come close to reproduce 
the very deep theoretical $I$ profiles. 
We have also carried out tests with the use of different micro- and macro-turbulent 
velocities, and we found that the choice of turbulent velocity does not significantly 
affect the rest intensity of the Stokes $I$ profiles.

From Figures~\ref{profiles} and \ref{clv-new} we notice that for $\mu \leqslant 0.35$ 
the observed line center $Q/I$ is less than the theoretical value and for 
$\mu > 0.35$ it is greater than the $Q/I$ predicted theoretically.  
Such a discrepancy was also encountered by \citet{mf94} while modeling the 
CLV of the line center $Q/I$ of the Ca~{\sc i}~4227~\AA\ line. 
The author found that the ratio of $(Q/I)_{\rm obs}$ and $(Q/I)_{\rm theory}$
at the line center was close to unity for smaller $\mu$ values ($\mu < 0.4$) and 
much greater than one for larger $\mu$ values. 
However her treatment did not include the stray light corrections. 
We recall that the observed profiles in Figure~\ref{clv-new} are corrected for the 
stray light.  
In \citet{mf94} though an explanation of the physical mechanism behind this enhancement in 
polarization for larger $\mu$ values was anticipated based on accelerated motions in the chromosphere, 
but it was not completely justified.

To examine this discrepancy further, we plot the variation exhibited by 
$Q/I$ along the spectrograph slit at the line center (solid line) and compare it with the $Q/I$ 
in the blue wings (dotted line) in Figure~\ref{slit-variation}. The observed 
line center $Q/I$ values are smoothed over a rectangular box corresponding to 
5$\arcsec$ to reduce contribution from noise. 
These smoothed values of the observed $Q/I$ at the line center are used for all 
further computations. 
From Figure~\ref{slit-variation} we see that 
the $Q/I$ at the line center shows more variation along the slit than at the blue wing peak.
This indicates the presence of varying horizontal magnetic fields and their possible role 
in modifying the line center $Q/I$. These varying magnetic fields can in turn 
be used to understand the observed line center $Q/I$ which are greater than the 
theoretically predicted values. One possible explanation for this discrepancy could be 
that observed line center $Q/I$ 
for $\mu > 0.35$ is enhanced due to the Hanle effect by these varying fields \citep[see also][]{mf94}. 
This enhancement is very prominent in case of
the near disk center observations \citep[see][]{FS11}. However, in our case, it sets in for 
$\mu > 0.35$ and 
increases as $\mu \to 1$. 
This is a clear evidence for highly structured, resolved, oriented magnetic 
fields (predominantly horizontal) in the solar atmosphere. 
However the dotted line in Figure~\ref{slit-variation} (for the blue wing) 
do not exhibit the type of spatial fluctuations that is seen for the line 
center (solid line). This is because the Hanle effect is absent in the wings. 
We also note that the spatial variation close to the limb in the blue wing is not 
really spectrally flat. The details regarding this will be 
discussed in Section~\ref{depth-dependence}. 

\subsubsection{The impact of temperature structure modifications on the 
standard model atmospheres}
In the previous section we described the necessity of constructing a new model in order to obtain 
a fit to the CLV of the Stokes profiles. 
To this end, a new model was constructed by combining two standard models after modifying 
their temperature structures at the desired heights. The new combined model thus constructed 
will provide a fit to the CLV of the observed $Q/I$. The 
physical consistency of the newly constructed atmospheric model with the modified temperature 
structure has been checked by verifying that it satisfies hydrostatic equilibrium at all 
heights.

Next we examine the fit to the CLV of the continuum intensity over a wavelength range spanning 
from the visible to the infrared. The theoretical continuum intensity obtained using the new model 
should fit the observed data 
at all the limb distances and for a range of wavelengths. 
Figure~\ref{clv-continuum} shows the limb darkening function computed using 
the standard models and our new model 
atmosphere $\overline{{\rm FALA}}$ + FALX
for a range of wavelengths and $\mu$ values. The theoretical values from different models 
are compared with
the observed data from \citet{nl94}. 
The dash-triple-dotted line represents the theoretical values from the new model 
$\overline{{\rm FALA}}$ + FALX. We see that the best fit to the observations is 
provided by the FAL-C model. The combined model though is successful in providing a CLV fit to the 
observed $Q/I$ and satisfies the equilibrium conditions, it does not provide the best fit to 
the observed CLV of the limb-darkening function and to the observed CLV intensity. 

This leads 
us to the conclusion that  it is indeed not possible to obtain a simultaneous fit to all 
the various types of data with a single 1-D model atmosphere. One needs a different atmosphere for 
each observable. 
In search of a single model which satisfies all the observational constraints, 
the next obvious step would be to use the two-component 
modeling approach with appropriate mixing ratios as done in \citet{holandsten07}. In the section below 
we discuss why we cannot adopt such a procedure in modeling the CLV of the Ca~{\sc i}~4227~\AA\ line.

\subsubsection{The two-component modeling approach}
\label{comparison}
In modeling the CLV observations of the Ca~{\sc ii} K line, 
\cite{holandsten07} explored the possibility of constructing a
two-component model atmosphere. This was constructed by 
mixing the results obtained from two standard model atmospheres
in appropriate ratios. Such a method was adopted by 
making CLV plots of the Ca~{\sc ii} K line as shown in their Figure~1. 
As seen from their figure, the original models FAL-X and FAL-C produce theoretical 
CLV curves which fall respectively above and below the observed CLV curve of $Q/I$. 
Hence they combine results from these two standard models with appropriate mixing ratios to 
achieve the required fit. 
In Figure~\ref{clv-plots} of the present paper we make similar plots 
of the CLV for the Ca~{\sc i}~4227~\AA\ line. 
As seen from Figure~\ref{clv-plots}, none of the standard model atmospheres 
produce a theoretical CLV curve which falls above the observed CLV curve. 
This does not allow us to apply the same kind of 
two-component modeling procedure as described 
by \citet{holandsten07}. This suggests that we need to go beyond 1-D modeling 
in the direction of 2-D or 3-D 
modeling to obtain a simultaneous fit to the ($I, Q/I$) at all the 
limb distances. Such efforts are beyond the scope of this paper. However 1-D 
models with modified temperature structures serve as a good initial step to 
such elaborate computations. The failure of 1-D modeling approach does not preclude the 
use of a given line profile for purposes like magnetic field determination. 
To demonstrate this fact, in the next section, we perform observational analysis 
of the Ca {\sc i} 4227 \AA{} line to 
determine the field strengths for smaller $\mu$.

\begin{figure*}
\centering
\includegraphics[scale=0.47]{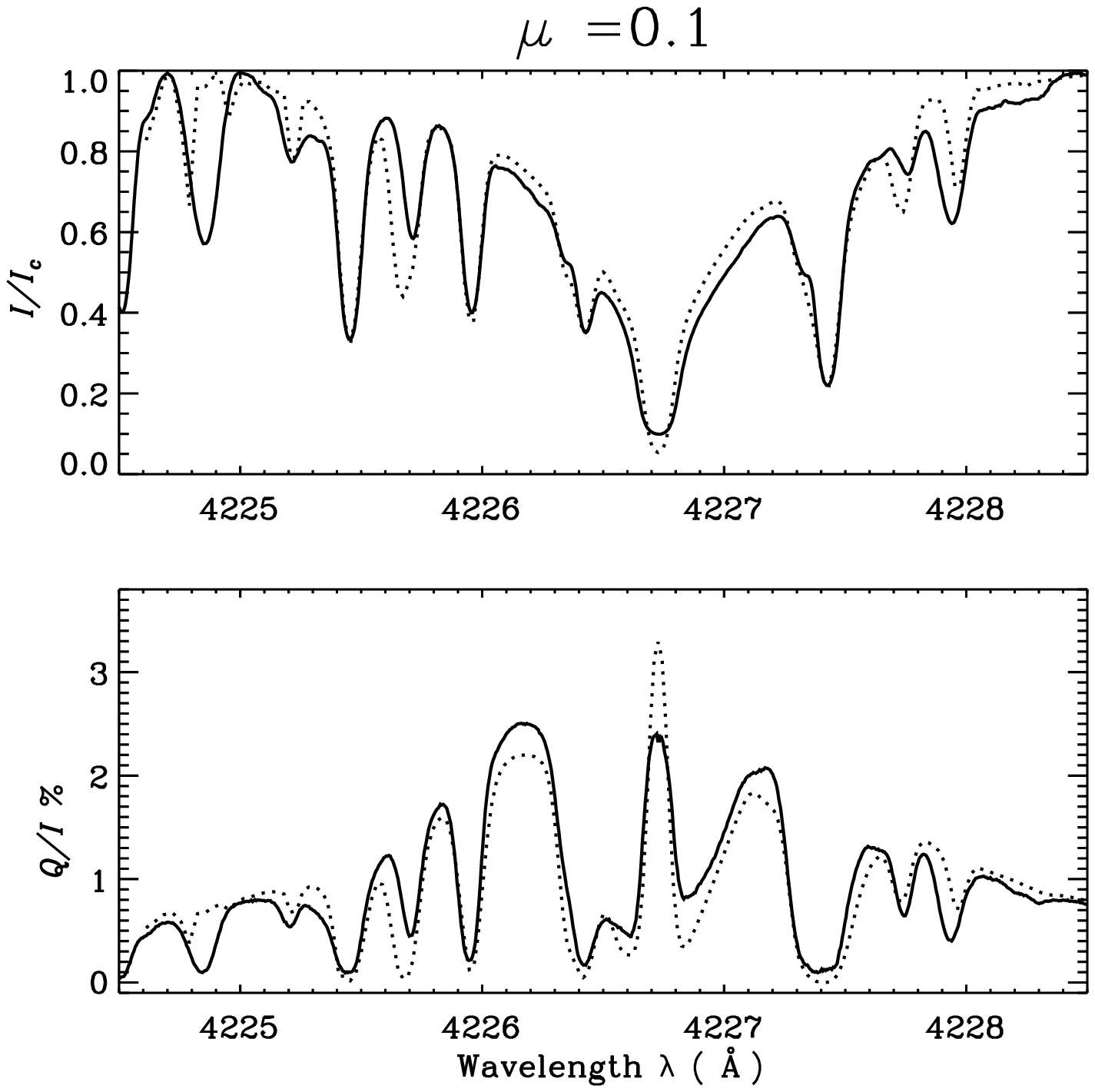}
\includegraphics[scale=0.47]{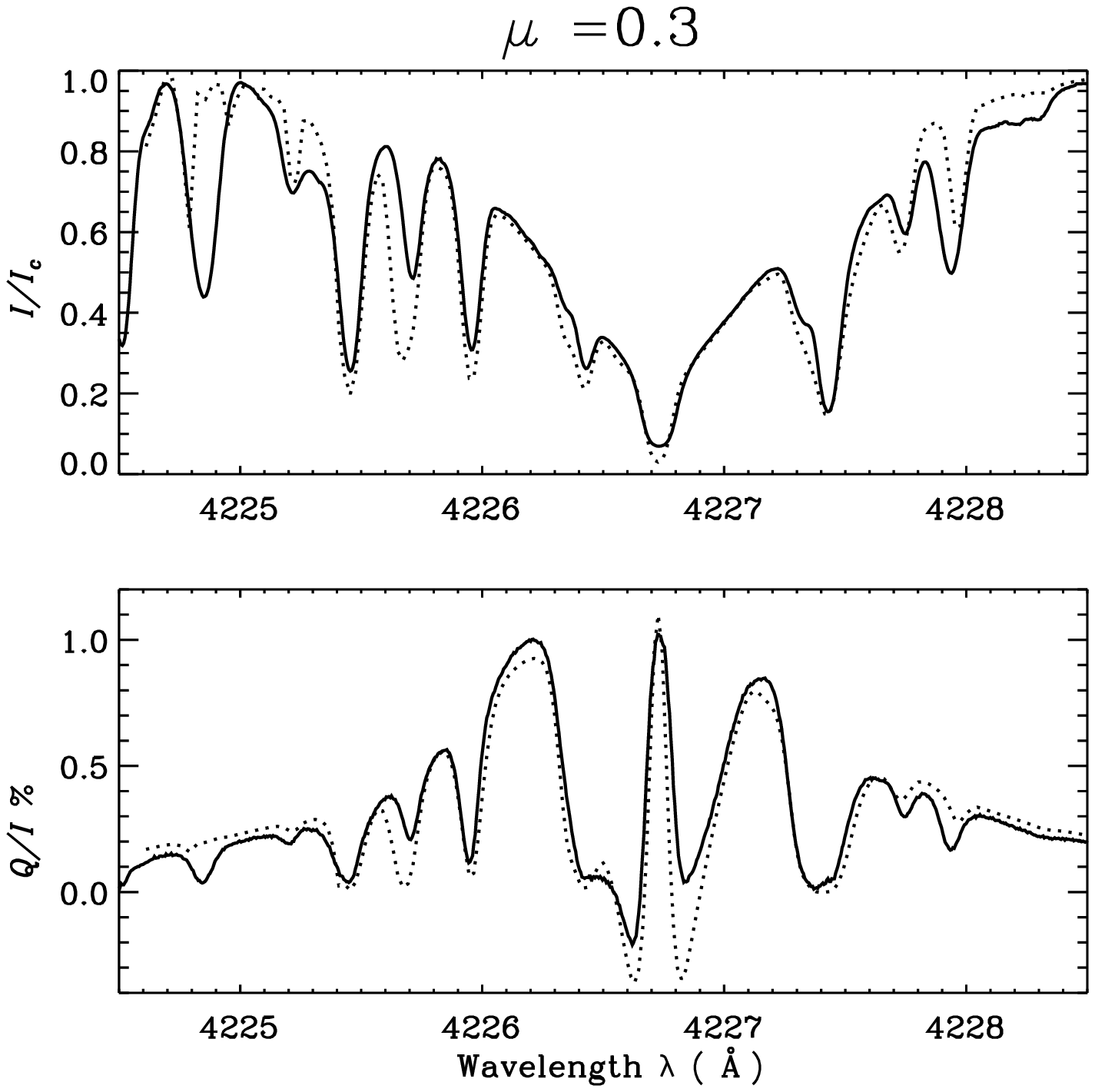}
\includegraphics[scale=0.47]{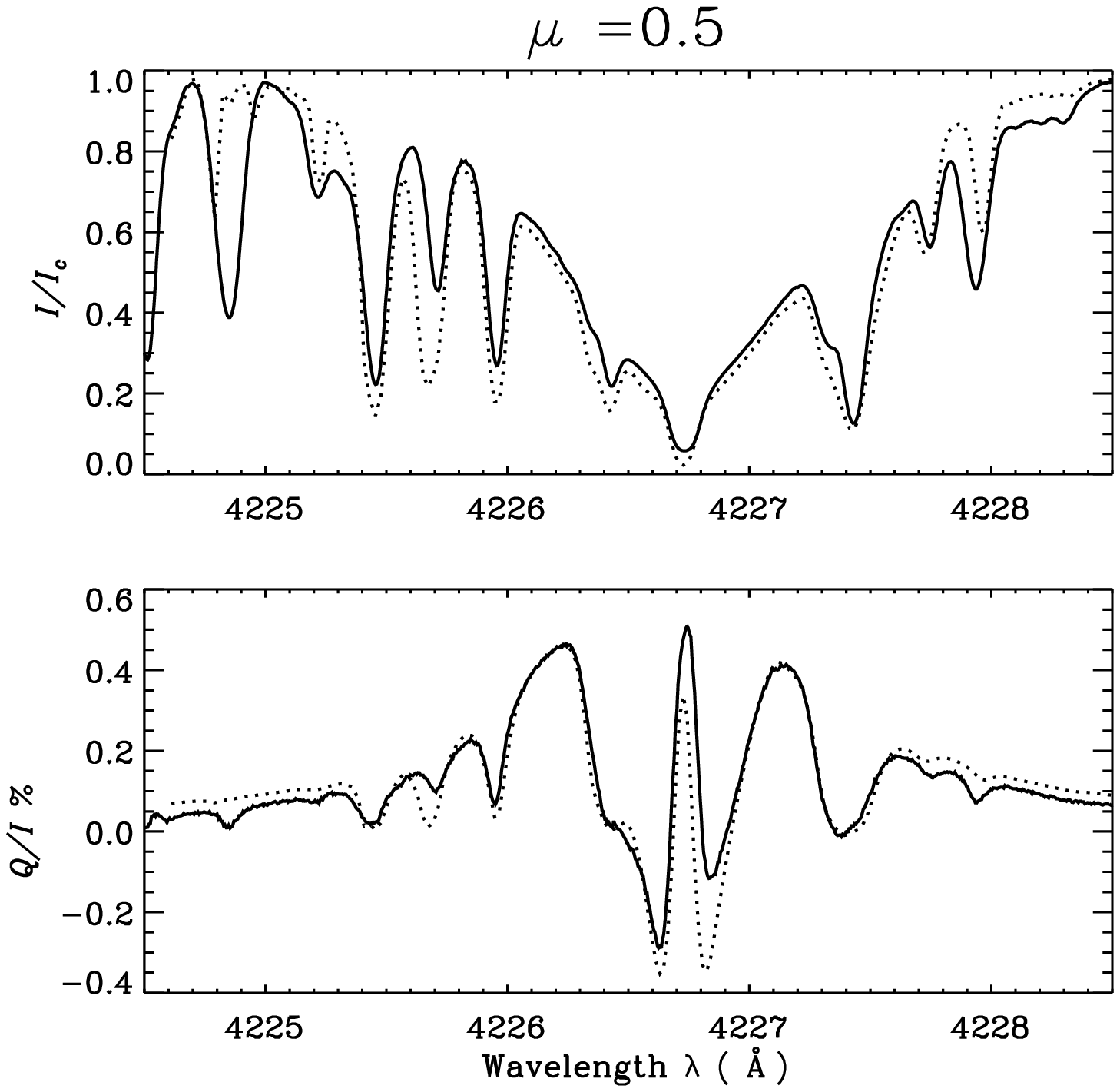}
\includegraphics[scale=0.47]{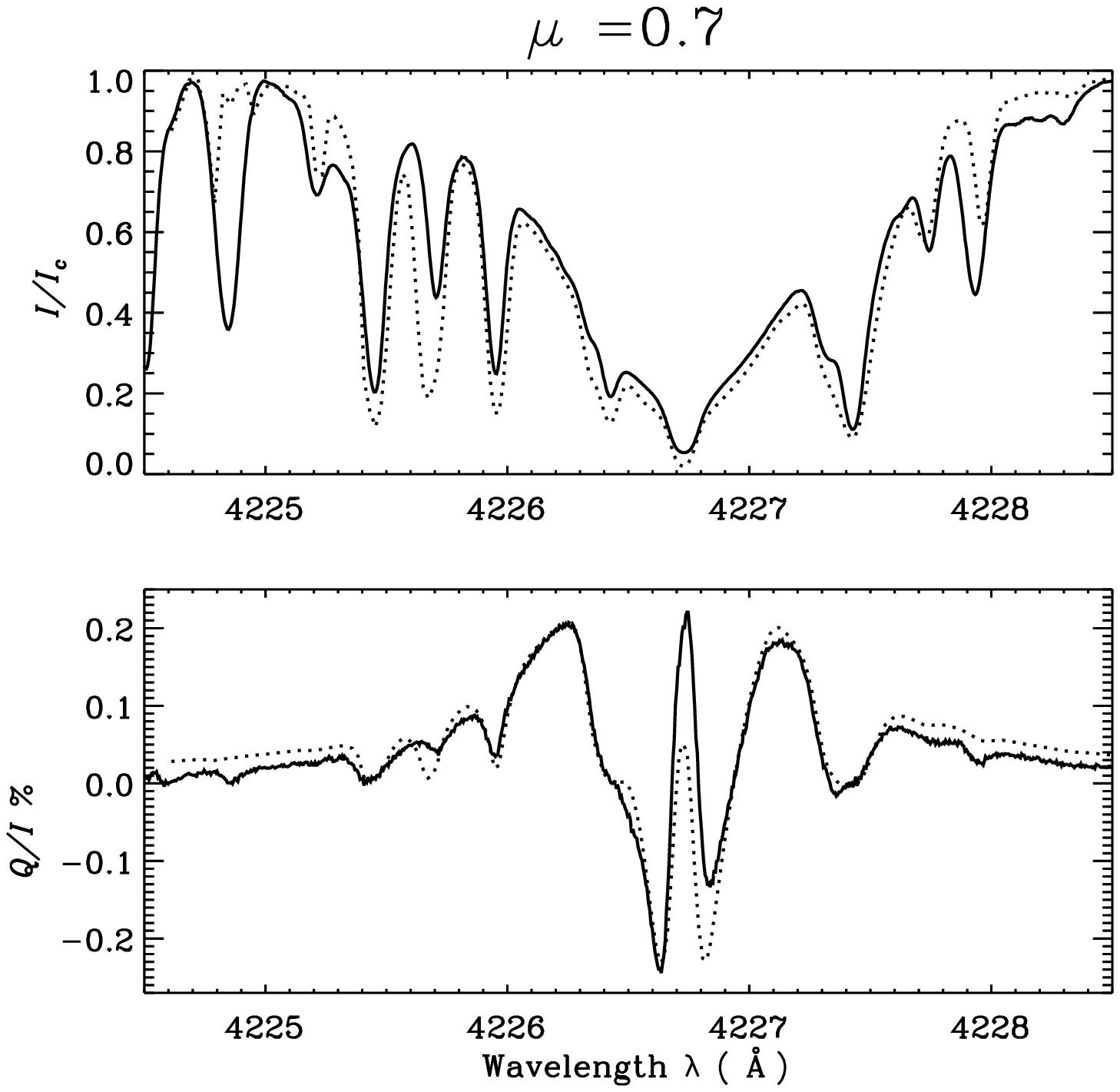}
\includegraphics[scale=0.47]{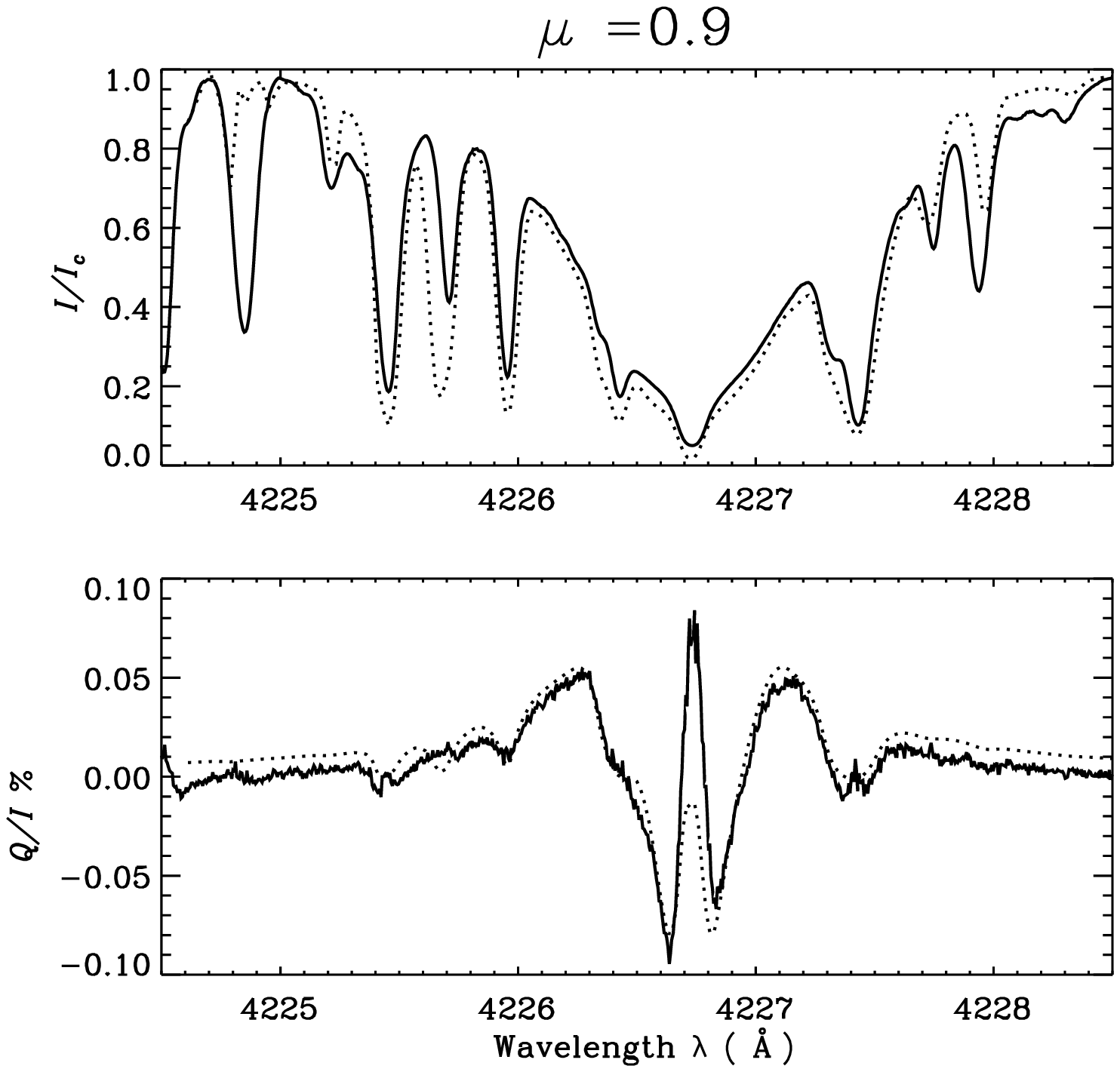}
\caption{Comparison between the observed (solid line) and the theoretical (dotted line) 
Stokes profiles $(I,Q/I)$ at different limb distances. The combined model atmosphere  
$\overline{{\rm FALA}} +{{\rm FALX}}$ is used to compute the theoretical profiles.}
\label{profiles}
\end{figure*}

\begin{figure*}
\centering
\includegraphics[height=6cm, width=16cm]{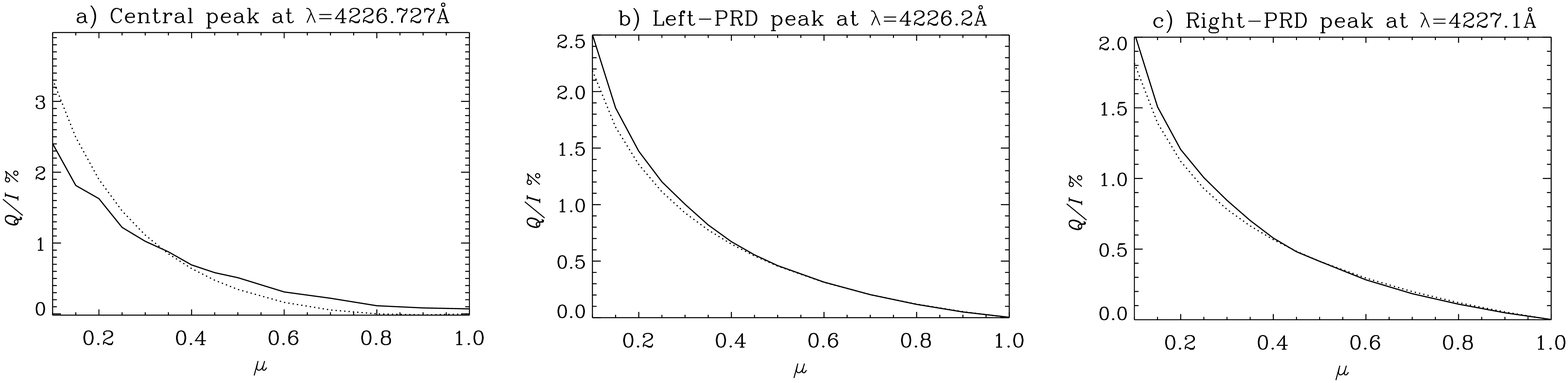}
\caption{Center to limb variation of linear polarization at three chosen wavelength positions. 
The model atmosphere used to obtain the theoretical profiles (dotted line) is the new combined model 
$\overline{{\rm FALA}}$ + ${{\rm FALX}}$.}
\label{clv-new}
\end{figure*}

\begin{figure*}
\centering
\includegraphics[scale=0.9]{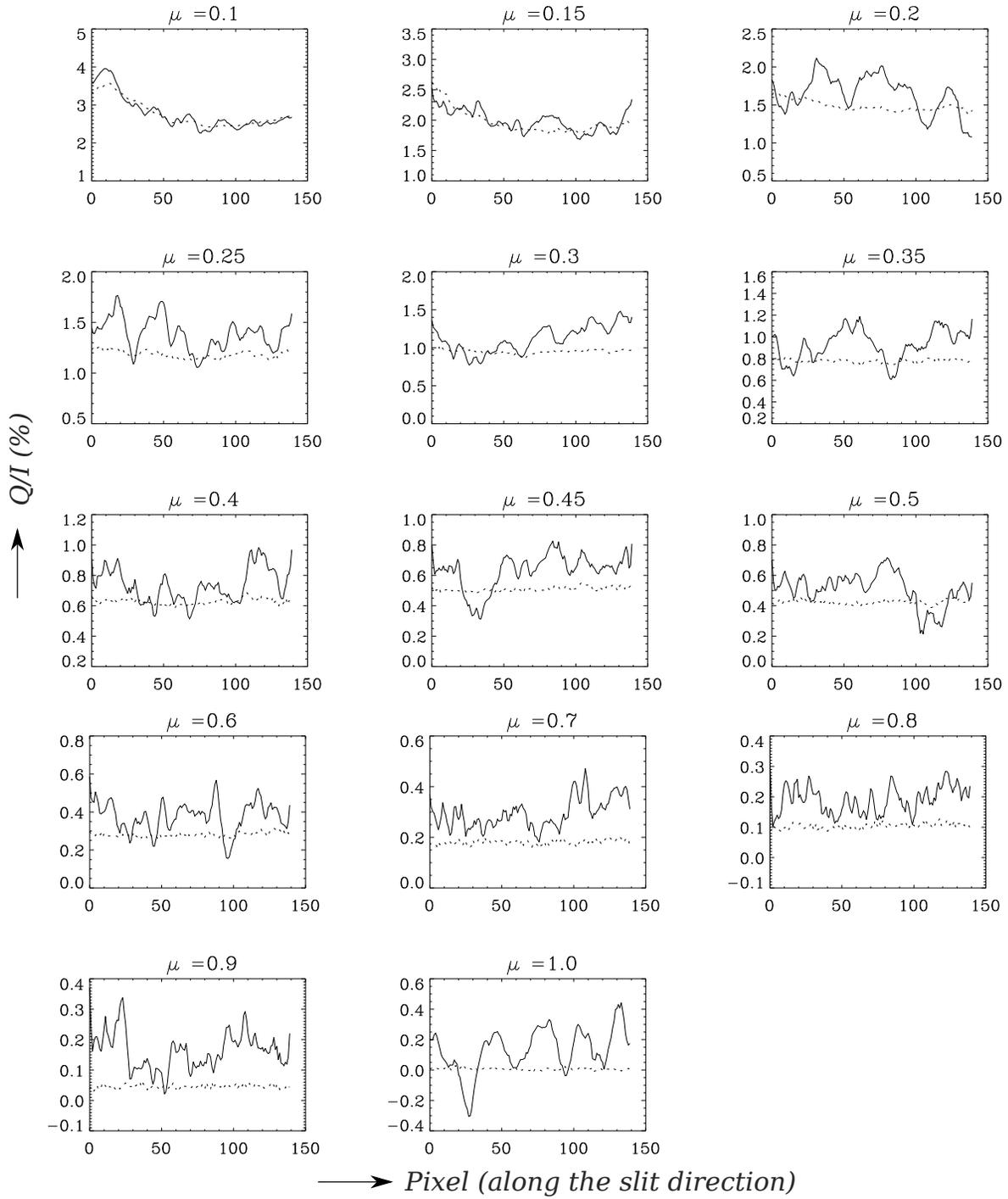}
\caption{Variation of $Q/I$ values along the slit for each $\mu$ position (marked over the plots). 
The solid and the dotted lines correspond to the $Q/I$ value at the line center and at the blue wing peak 
respectively.}
\label{slit-variation}
\end{figure*}

\begin{figure*}
\centering
\includegraphics[scale=0.7]{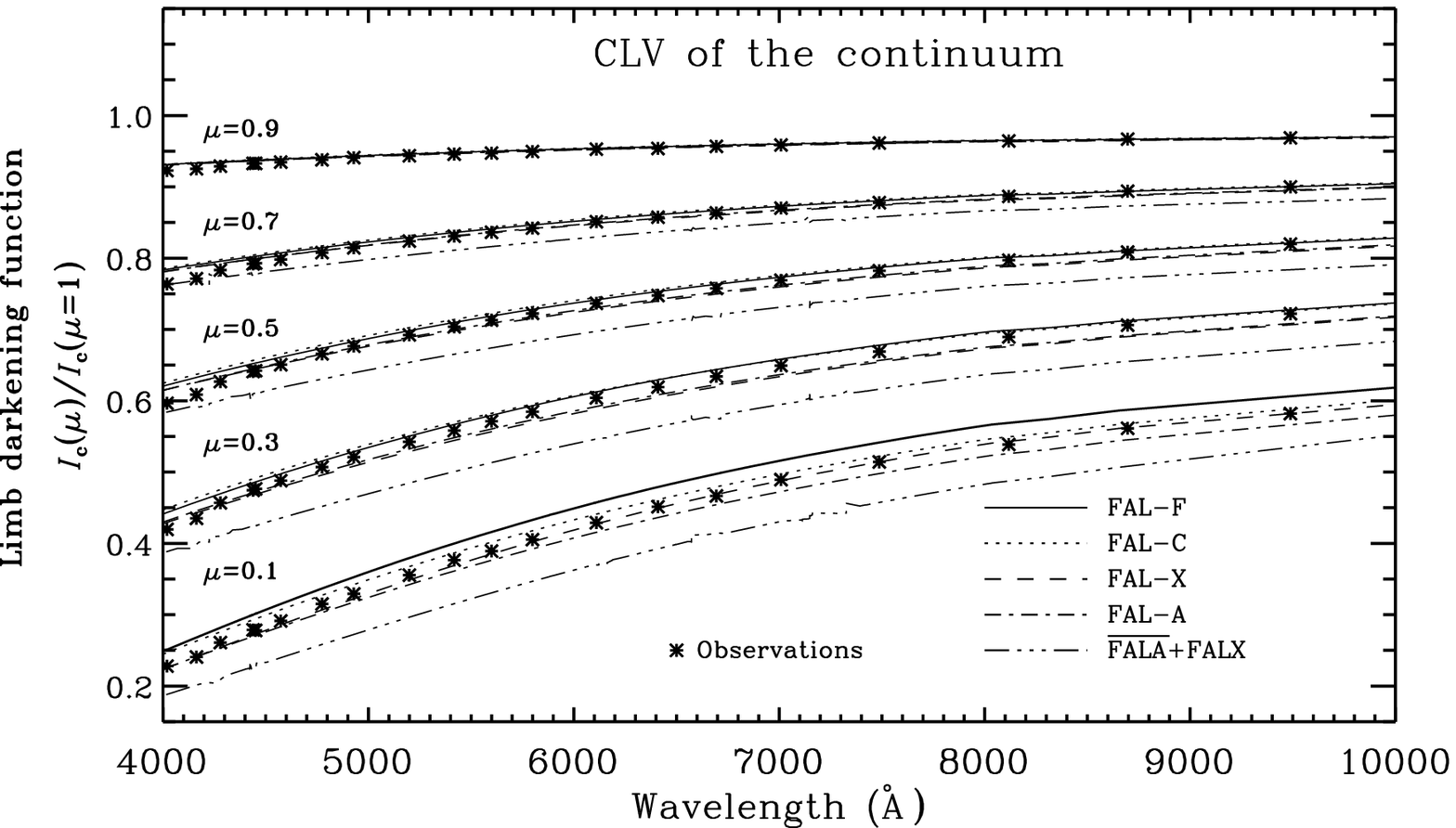} 
\caption{Comparison between the  observed data for the CLV of continuum intensity
from \citet{nl94} for a range of wavelengths  and 
the theoretical values from different 
model atmospheres including $\overline{{\rm FALA}}$ + FALX. }
\label{clv-continuum}
\end{figure*}

\section{Determination of the field strength}
\label{depth-dependence} 
In the present section we use an approach similar to that of \citet{bianda98,bianda99} to determine the 
field strength at different limb distances. Since the observed $Q/I$ is influenced by so many factors 
besides the magnetic field, it is imperative to apply {\it differential} techniques to isolate the 
Hanle effect from the multitude of other effects. This can be done by using the wing polarization as 
a reference, since it has been well established that the Hanle effect only operates in the line core 
but is absent in the wings. 

If we in Figure~\ref{slit-variation} compare $Q/I$ in the line core (solid lines) and in the blue wing 
(dotted lines), we notice that the line core exhibits large spatial variations along the slit, 
in contrast to the blue wing. Nevertheless, the blue wing polarization exhibits large-scale slow drifts 
along the slit, which increase significantly as we approach the limb. Much of this can be explained in 
terms of a geometric effect due to the limb curvature. Since the solar limb is curved while the slit 
is straight, the limb distance (or $\mu$) will vary along the slit. This effect will increase in 
significance as we get closer to the limb. It is an effect that is nearly identical for the line core 
and wing (since the core and wing have nearly the same relative center-to-limb variations) and 
therefore can be eliminated when forming the core to wing ratio. Similarly, any other unidentified 
instrumental effect would ratio out. In principle there may also be non-magnetic effects of solar 
origin, like spatial variations of the radiation-field anisotropy, which may be different between core 
and wings and therefore would not fully ratio out (although they should be suppressed when forming the 
ratio, since the non-magnetic fluctuations in the core and wings are not uncorrelated). However, 
with our rather low spatial resolution and long integration times, these solar effects are expected to 
be miniscule. 

We therefore have strong reasons to believe that practically all the spatial fluctuations that we 
see in the $Q/I$ core to wing ratio is exclusively due to magnetic fields via the Hanle effect. 
Instead of directly using this ratio as our differential measure, we can scale it with the slit 
average of the wing polarization, to express it in polarization units. This scaling is equivalent to 
the assumption that the wing polarization should be spatially flat after all effects of limb curvature, 
unidentified instrumental effects, and solar non-magnetic effects have been corrected for. We thus 
correct the line core polarization amplitude with the following relation
\begin{equation}
 (Q/I)_{corrected}^{line\ center} = \frac{(Q/I)_{uncorrected}^{line\ center}}{P_b}\ {<P_b>},
 \label{correction}
\end{equation}
where $P_b = (Q/I)^{blue\ wing\ peak}$ for each pixel and $<P_b>$ is the spatial 
average of $P_b$ along the slit. With this correction we plot in Figure~\ref{clv-slit} the CLV of 
the spatially averaged $Q/I$ at the blue wing and the corrected $Q/I$ at the line center. 
Each ``plus'' symbol in bottom panel of Figure~\ref{clv-slit} represents the value of $Q/I$ 
at each pixel corresponding to the line center. We notice large spatial variations along the 
slit in the corrected $Q/I$ line-center data. This effect is 
exclusively due to the magnetic fields via the Hanle effect. In order to find the field strengths 
that contribute to such spatial variation, we follow the method used in \citet{bianda98,bianda99}. 
We would like to note that in both these papers, for the data analysis, 
the authors use observations taken at different periods. However in our analysis we 
consider only one single set of observations and the variation of $Q/I$ along the slit 
in these observations. 
To this end we construct the envelopes (continuous lines in Figure~\ref{clv-slit}) 
to our data set, using the analytical relation 
\begin{equation}
 \frac{Q}{I} = \frac{a (1-\mu^2)}{\mu+b}.
\label{empirical}
\end{equation} 
This relation was first introduced by \citet{jos97} where $a$ and $b$ are the best fit free parameters. 
For our studies we have chosen the same set of free parameters as given in \citet{bianda98,bianda99}. 
From top panel of Figure~\ref{clv-slit} we see that the dashed line ($a$ = 0.33\%, $b$ = 0.02) gives a 
good fit to the spatially averaged observed CLV profile in the blue wing. 
In the bottom panel of Figure~\ref{clv-slit} 
we use three different set of free parameters $a$ and $b$ to construct envelopes for 
the line center data. 
The envelopes constructed using the analytical relation given in Equation~(\ref{empirical}) 
represent the `non-magnetic value' 
and all the values lying below this envelope are considered as the depolarized $Q/I$  
values due to the Hanle effect. From our modeling efforts we know that the 
magnetic fields cause an enhancement in the polarization value for $\mu > 0.35$ 
\citep[also see][]{mf94}. Hence this envelope 
fitting method is good for $\mu \leqslant 0.35$ and becomes questionable for $\mu$ larger than about 
0.35. 
However the transition between the large angle scattering 
and small angle scattering is gradual and smooth. For large $\mu$ we gradually enter into the regime of 
forward-scattering Hanle effect, for which the kind of techniques developed by \citet{FS11} have to be adopted 
to derive the field strengths. Full radiative transfer modeling is naturally needed for 
intermediate $\mu$ values. 
Only for smaller $\mu$ values it is possible to use a method that avoids the need for 
radiative transfer. In this approach, we  first extract an observed depolarization factor via the 
envelope method and then convert 
this depolarization into field strength.

Thus we first determine the ratio between the line center $Q/I$ and the corresponding 
envelope value. 
This ratio represents the depolarization factor caused by the Hanle effect for each pixel. 
The conversion of this factor into field strength is dependent on the choice of the envelope, 
since it represents a 
single observable, while the magnetic field vector is characterized by three parameters (its spatial 
components). The magnetic field is therefore underdetermined, so a conversion cannot be unique, 
but it is still meaningful 
in a statistical sense, as explained in the following.

For photospheric spectral lines an interpretational model with a spatially averaged micro-turbulent 
field distribution could be used to convert Hanle depolarization into field strength  
\citep[][]{sten82, jos94}, because the absence of $U/I$ polarization in combination with insignificant 
spatial variations (at resolved scales) in $Q/I$ made such a micro-turbulent interpretational model 
unavoidable \citep[see also][]{jos13}. The situation is however entirely different for strong 
chromospheric lines like the Ca {\sc i} 4227\,\AA\ line, which always exhibits large spatial line-core variations 
in both $Q/I$ and $U/I$, like those in Figure~\ref{slit-variation}. As we have spatially resolved these 
variations, they should ideally be interpreted in terms of resolved, oriented fields rather than 
angular distributions. However, since the field vector is underdetermined by the single depolarization 
factor, we need to eliminate the ambiguity by using the statistical approach, dealing with each 
depolarization factor as if it were obtained through averaging over an ensemble of field elements. 
This approach will give field strength values that are meaningful as averages in a statistical sense.

Vertical fields are immune to the Hanle effect, depolarization can only occur if the field has a 
substantial inclination with respect to the vertical direction. Horizontal fields give the largest 
depolarization. If we assume the fields to be horizontal, but with orientations that are random in 
azimuth angle, and let the Hanle depolarization be determined by an ensemble average over such a field 
distribution, then the field strengths that we extract from this model can be considered to represent 
{\it lower limits} to the true average field strength (since there may exist less inclined fields 
that are less ``visible'' to the Hanle effect).

Chromospheric fields are expected to be largely horizontal, forming a ``canopy'' over the 
underlying photosphere. The Hanle depolarization factor $k_H$ for a horizontal field distribution 
with random azimuths can be written as \citep[][]{sten82, jos94} 
\begin{equation}
 k_H = 1 - 0.75\sin^2 \alpha_2,
 \label{depol}
\end{equation}
where the Hanle mixing angle $\alpha_K$ is given by
\begin{equation}
 \tan \alpha_K = \frac{K B}{B_0/k_c^{(K)}}.
\label{mag}
\end{equation}
$B$  is the field strength to be determined, $K$ = 1 or 2, 
$k_c^{(K)}$ is the collisional branching ratio for the 2$K$-multipole and $B_0$ is the 
characteristic field strength for the Hanle effect.\\

Figure~\ref{histogram1} shows the histograms of the field strengths obtained when using 
Eqs.~(\ref{depol}) and (\ref{mag}) for all pixels along the spatial direction. 
By definition, a depolarization factor should be less than or equal to unity, otherwise 
it is unphysical. However, some points give an unphysical depolarization factor, both because 
there is scatter of the $Q/I$ values due to measurement noise, and because the chosen 
envelope may be too low. In such cases the field strength used for the histograms in 
Figure~\ref{histogram1} is set to zero. The number of such zero field points depends on the 
choice of  envelope and increases as we move away from the limb because of the increasing 
contribution from the forward-scattering Hanle effect.

Figure~\ref{histogram1} shows how the field-strength fluctuations along the slit vary 
with different limb distance. For the derivation of the CLV of the average field strength, 
which is shown in Figure~\ref{field}, we do not average the field-strength histograms of 
Figure~\ref{histogram1}, because they are affected by measurement noise in a non-linear way 
(including the truncation used for the unphysical values), but we instead average the measured 
$Q/I$ along the slit (causing the Gaussian instrumental noise to get greatly suppressed), and 
then convert the average $Q/I$ to field strength. Since the height of line formation increases 
with decreasing $\mu$, the $\mu$ variation displayed by Figure~\ref{field} may be interpreted in 
terms of a height variation of the field. In view of the limited statistical sample and the crudeness 
of the interpretational model the minor variations with $\mu$ in Figure~\ref{field}, which 
are similar to the ones obtained by \citet{bianda98,bianda99}, are not significant but are 
compatible with approximate constancy of the average field strength over the height range covered 
by our $\mu$ range.

Note that we have limited the $\mu$ range in Figure~\ref{field} to 0.1 - 0.35, because as 
mentioned before the envelope method is not applicable for larger $\mu$ values. 
Note also how the derived mean field strength depends on the choice of envelope. 
In spite of these uncertainties, the values are generally limited to the range 6 - 10\,G. 
We cannot choose envelopes significantly lower than the one represented by the 
solid line (in the bottom panel of Figure~\ref{slit-variation}), 
because one would then get an excessive number of unphysical depolarization factors. 
Therefore the 6\,G value can be seen as representing a kind of lower limit for the average 
field strength. 

\begin{figure*}
 \centering
 \includegraphics[scale=0.5]{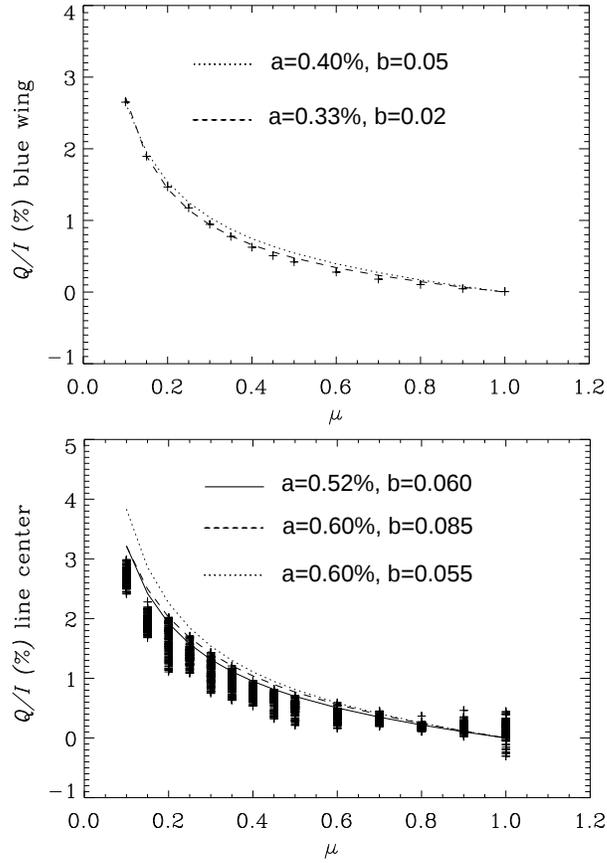}
 \caption{CLV of $Q/I$ at the blue wing PRD peak (top panel) and at the line center (bottom panel). 
Plus sign represents the $Q/I$ value at each pixel along the slit in the bottom panel and 
spatially averaged value of $Q/I$ in the top panel. 
The solid, dotted and dashed curves are obtained 
using the empirical relation given in Equation~(\ref{empirical}). The corresponding value of the free 
parameters $a$ and $b$ are indicated in the figure.}
  \label{clv-slit}
\end{figure*}

\begin{figure*}
 \centering 
 \includegraphics[height=18.0cm,width=14cm]{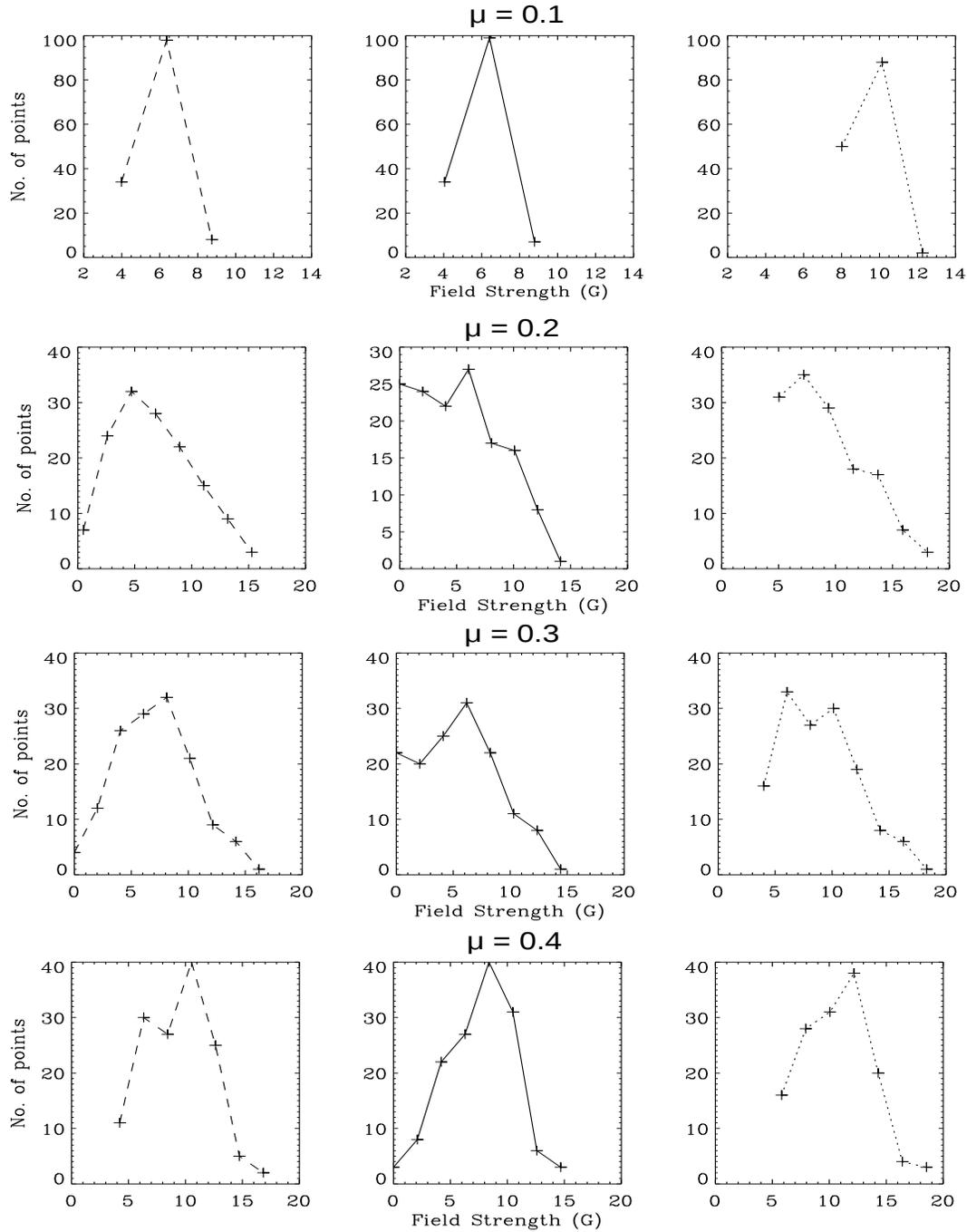}
 \caption{Histogram of the field strength at different $\mu$ positions. Field 
strengths are computed for each depolarization value in the spatial direction. Different 
panels along the row for each $\mu$ correspond to field strengths obtained using different envelopes. 
The solid, dotted and dashed lines respectively correspond to the solid, 
dotted and dashed envelopes in Figure~\ref{clv-slit}.} 
 \label{histogram1}
\end{figure*}

\begin{figure*}
 \centering
 \includegraphics[scale=0.6]{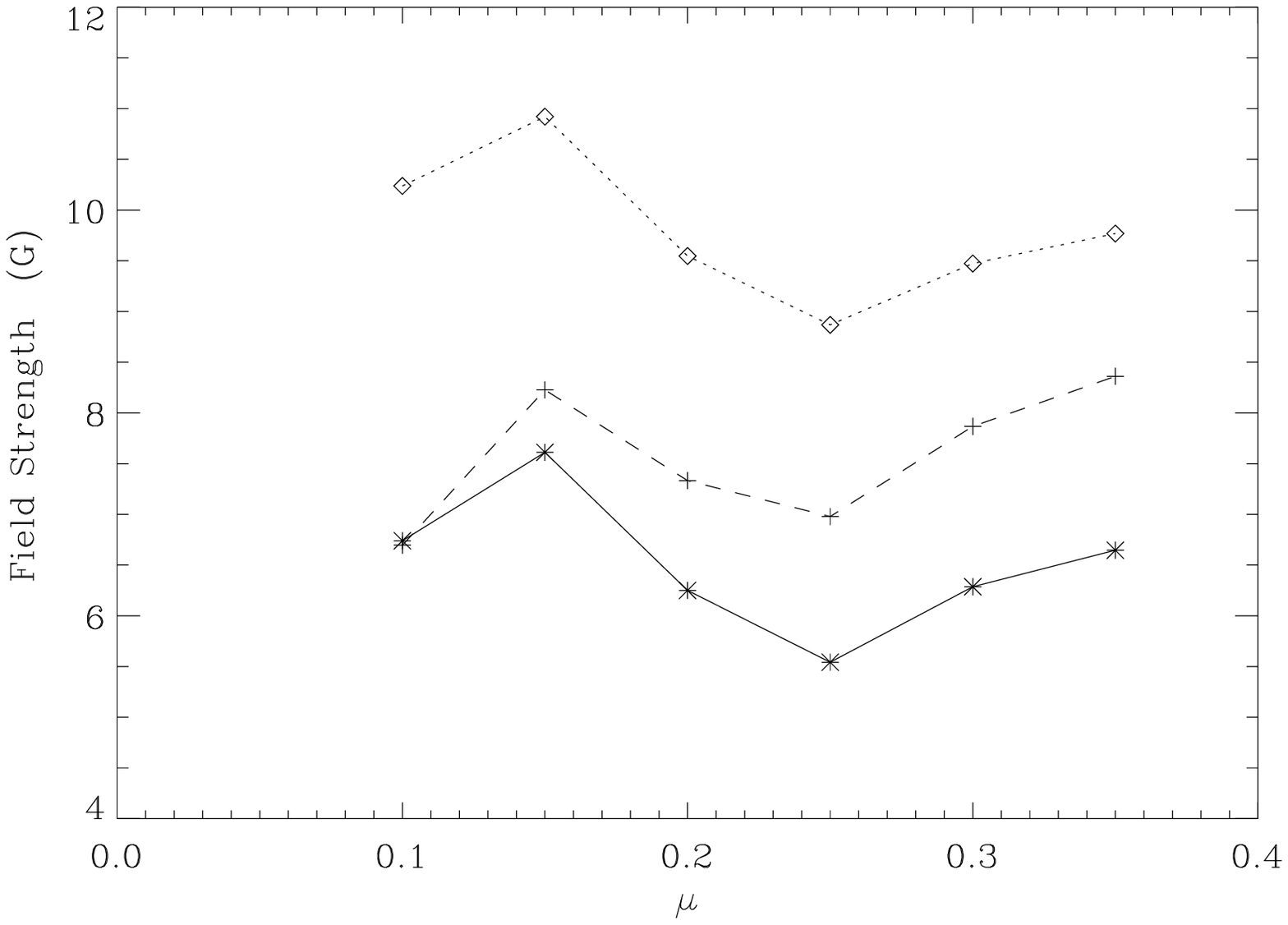}
 \caption{Mean value of field strength (G) derived from averaged $Q/I$ value along the slit. 
The solid, dotted and dashed lines correspond to the mean field strength derived from the 
corresponding envelopes indicated in Figure~\ref{clv-slit}. 
}
\label{field}
\end{figure*}

\section{Conclusions}
\label{conclusion}
To understand the depth dependence of various physical quantities in the Sun, 
like the magnetic fields, it is important to model 
the CLV observations of suitable atomic and molecular lines. 
In this paper we have attempted to model such CLV observations of the well 
known Ca {\sc i} 4227 \AA {} 
line. In our approach we take into account the effects of
partial frequency redistribution (PRD) and radiative transfer. 
The observations of this line were carried out in quiet regions 
on the Sun at 14 positions starting 
from the limb up to the disk center on October 16, 2012 at IRSOL in Switzerland.
This line has the largest degree of linear polarization in the visible region of the 
Second Solar Spectrum
and can be modeled by considering a simple two-level atom picture. 
When trying to model this CLV data we find that none of the standard atmospheric models, 
attempted by us,
like the FAL-F, FAL-A, FAL-C, and FAL-X could simultaneously fit 
the observed $(I,Q/I)$ profiles at all the limb distances. 
To model the CLV of the line center intensity we need FAL-F model which is the 
hottest and to 
model the CLV of the linear polarization at line center we need the coolest model FAL-X. 
In order to obtain a fit to the observed Stokes profiles, modifications in the temperature structure 
of the standard models become necessary. With suitable modifications in the desired height range,
we constructed $\overline{\rm FALA}$ and later combined it with FAL-X.
While the $\overline{\rm FALA}$ model gives a good fit to the PRD peaks, the 
FAL-X gives a good fit to the line center. The combined model has 
the temperature structure of 
$\overline{\rm FALA}$ up to 400 km, and that of 
FAL-X in the upper layers. This new combined model atmosphere
gives a good fit to the entire $Q/I$ at all values of $\mu$. 
Also in modeling efforts we found that the Hanle 
effect not only depolarizes the line core of $Q/I$ 
(which is true for smaller $\mu$'s) 
but also enhances the line core $Q/I$ for larger $\mu$ values. This might 
be due to the highly structured horizontal magnetic fields in the solar atmosphere.  

Though the new combined model provides a fit to the CLV of the observed $Q/I$, 
it fails to reproduce the 
observed CLV of the continuum limb-darkening function and the CLV of the observed 
line core intensity. 
This failure of 1-D models in order to simultaneously fit the observed 
$(I,Q/I)$ CLV profiles do not restrain the use of the Ca {\sc i} 4227 \AA{} as 
a tool to map the magnetic fields. To support this claim, 
we carried out observational 
analysis to determine field strength using 
the Ca {\sc i} 4227 \AA{} for smaller $\mu$ values. 
  
To conclude, it appears that no single 1-D atmosphere can completely 
provide a good representation of the actual solar atmosphere. 
This shows that the solar atmosphere has a far more complex structure. 
To simultaneously satisfy the various observational constraints it is therefore 
unavoidable to go beyond such 1-D models - a difficult problem that needs to be approached step by step. 
This conclusion is not at all a technical failure 
meaning that our inability to obtain a simultaneous perfect fit to the CLV of the ($I$, $Q/I$) 
has nothing to do with the weakness of our approach or the method followed in using 
1-D solar atmospheres. Instead it is a ``profound failure'' indicating that the atmosphere of the 
Sun has such a complexity that it is not possible to represent 
it in terms of a single 1-D atmosphere. It could mean that the use of 1-D models 
for interpretations of the Second Solar Spectrum may give results that are 
physically incorrect (since they do not represent solar conditions), 
although the results may formally be mathematically correct. However, 1-D modeling efforts 
may still provide a guideline to the more systematic and sophisticated modeling efforts. 

\begin{acknowledgments}
We are grateful to  Dr. Han Uitenbroek for providing us with his
realistic atmospheric modeling code.
Research at IRSOL is financially supported by State Secretariat for
Education, Research and Innovation, SERI, Canton Ticino, the city of
Locarno, the local municipalities,
the Foundation  Aldo e Cele Dacc\`o  
and the Swiss National Science
Foundation grant 200021-138016. RR acknowledges financial support by the
Carlo e Albina Cavargna foundation. We are grateful to the referee for 
very detailed and useful comments which helped in improving the paper 
substantially.
\end{acknowledgments}

\end{document}